\documentclass[journal]{IEEEtran}
\usepackage{amsmath,amssymb,algorithm,algorithmic,graphicx}
\newtheorem{lem}{Lemma}

\newcommand{\figref}[1]{Figure~\ref{#1}}
\newcommand{\eg}{e.g.,~}
\newcommand{\ie}{i.e.,~}
\newcommand{\argmin}{\mathop{\rm argmin}}

\newcommand{\bfn}{\boldsymbol n}

\newcommand{\bfu}{\boldsymbol u}

\newcommand{\bfc}{\boldsymbol c}
\newcommand{\bfd}{\boldsymbol w}
\newcommand{\bfr}{\boldsymbol r}
\newcommand{\bfv}{\boldsymbol v}
\newcommand{\calF}{\mathcal F}
\newcommand{\calTF}{\mathcal{TF}}
\newcommand{\hatQ}{\hat Q}
\newcommand{\setU}{U}

\title{On Periodic Reference Tracking Using Batch-Mode Reinforcement Learning with Application to Gene Regulatory Network Control}
\author{Aivar Sootla, Natalja Strelkowa, Damien Ernst, Mauricio Barahona, Guy-Bart Stan \thanks{
Aivar Sootla and Guy-Bart Stan (corresponding author) are with the Centre for Synthetic Biology and Innovation and the Department of Bioengineering, Mauricio Barahona is with the Department of Mathematics, Imperial College London, UK {\tt\small \{a.sootla, g.stan, m.barahona\}@imperial.ac.uk}, Natalja Strelkowa is with Boehringer-Ingelheim Pharma GmbH \& Co KG., Germany, {\tt\small natalja.strelkowa@boehringer-ingelheim.com}, and Damien Ernst is with the Montefiore Institute, University of Li\`ege, Belgium {\tt\small dernst@ulg.ac.be}}}
\begin{document}
\maketitle
\begin{abstract}
In this paper, we consider the periodic reference tracking problem in the framework of batch-mode reinforcement learning, which studies methods for solving optimal control problems from the sole knowledge of a set of trajectories. In particular, we extend an existing batch-mode reinforcement learning algorithm, known as Fitted Q Iteration, to the periodic reference tracking problem. The presented periodic reference tracking algorithm explicitly exploits a priori knowledge of the future values of the reference trajectory and its periodicity. We discuss the properties of our approach and illustrate it on the problem of reference tracking for a synthetic biology gene regulatory network known as the generalised repressilator. This system can produce decaying but long-lived oscillations, which makes it an interesting system for the tracking problem. In our companion paper we also take a look at the regulation problem of the toggle switch system, where the main goal is to drive the system’s states to a specific 
bounded region in the state space.
\end{abstract}


\section{Introduction}
There are many problems in engineering, which require forcing the system to follow a desired periodic reference trajectory (\eg in repetitive control systems \cite{hara1988repetitive}). One approach to solve these problems is to define first a distance between the state of the system $\bfn_t$ and the desired reference point $\bfr_t$ at time $t$. Then explicitly seek for a control policy that minimises this distance for all $t$ subject to problem specific constraints. In the case of systems in the Euclidean state-space, for example, the Euclidean distance can be used.

In this paper, we are interested in solving reference tracking problems of discrete-time systems using an optimal control approach. In our setting, the dynamics of the system is only known in the form of one-step system transitions. A one-step system transition is a triplet $\{\bfn, \bfu, \bfn^+\}$, where $\bfn^+$ denotes a successor state of the system in state $\bfn$ subjected to input $\bfu$.
Inference of near-optimal policies from one-step system transitions is the central question of batch-mode reinforcement learning~\cite{fonteneau2012batch, busoniu2010reinforcement}. Note that reinforcement learning~\cite{Sutton1998} focuses on a more general problem of policy inference from interactions with the system. Usually, in batch-mode reinforcement learning, very few assumptions are made on the structure of the control problem. This gives batch-mode reinforcement learning algorithms a high degree of flexibility in comparison with other control methods. Batch-mode reinforcement learning has had numerous applications 
in many disciplines such as engineering~\cite{riedmiller2005neural}, HIV treatment design~\cite{Stan08}, and medicine~\cite{murphy2003optimal}. In this paper, one batch-mode reinforcement learning algorithm is considered, namely the Fitted $Q$ Iteration algorithm~\cite{Ernst2005c}. This algorithm focuses on the computation of a so called $Q$ function, which is used to compute a near-optimal control policy. The algorithm computes the $Q$ function using an iterative procedure based on the Bellman equation, a regression method and the collected samples of system trajectory. 

The focus of this paper is an extension of the Fitted $Q$ Iteration to the reference tracking problem. In our extension, we explicitly take into account the future reference trajectory values and the period $T$. Moreover, regression is separated into $T$ independent regression problems, which can save computational effort. The algorithm is implemented in Python using open-source packages and libraries \cite{pedregosa2011scikit},  \cite{joblib}, \cite{Hunter:2007}, \cite{scipy}.
\begin{figure}[t]
\centering
\includegraphics[width=0.7\columnwidth]{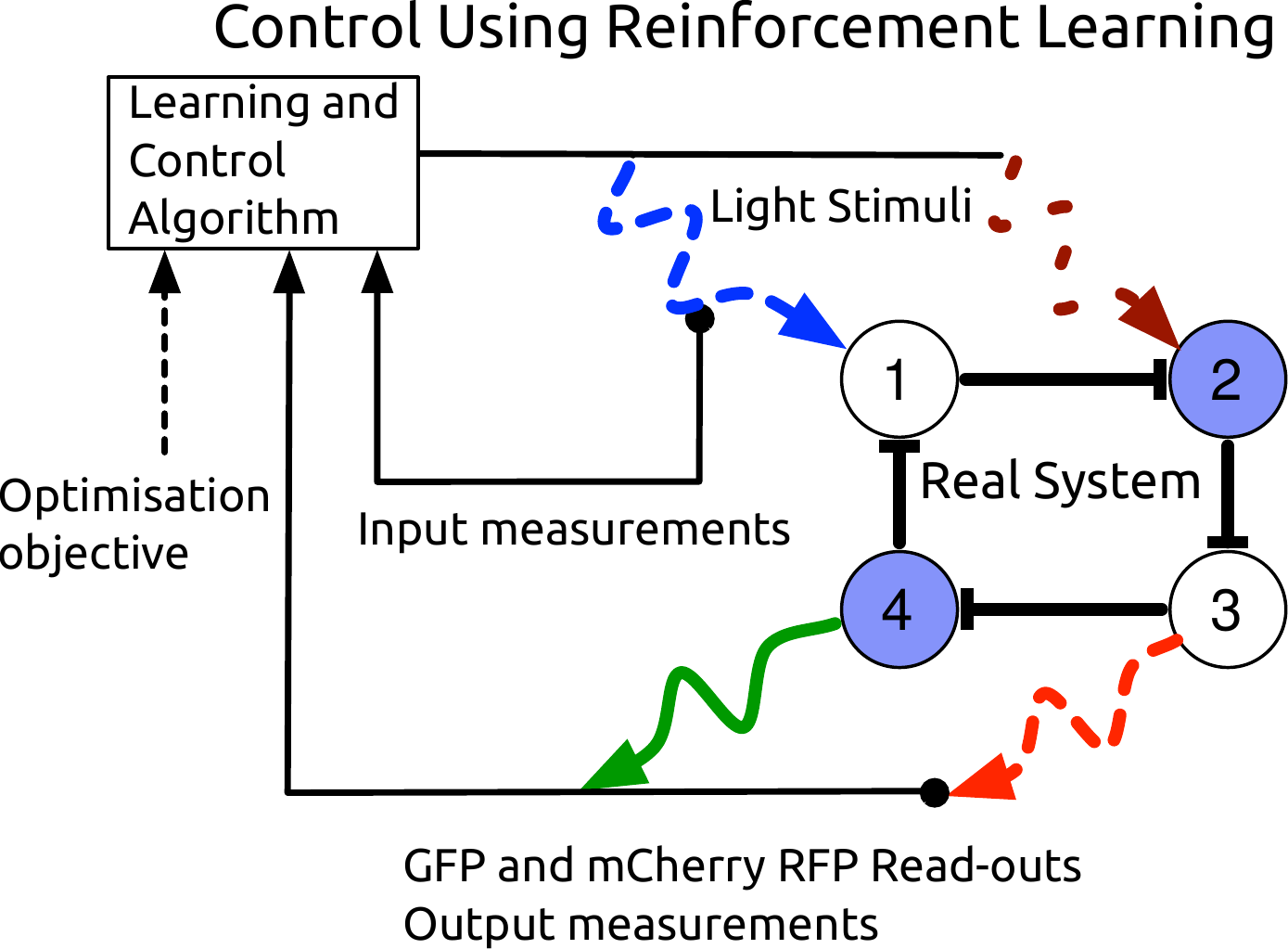}
\caption{A schematic depiction of an exogenously controlled generalised repressilator. Numbered circles represent different genes. An arrow with a flat end connecting two circles represents a repressive action from one gene product onto the other one. For example, gene $2$ is repressed by the protein product of gene $1$. } \label{fig:gr}
\end{figure}

To illustrate our reference tracking method, we consider its application to a synthetic biology gene regulatory network known as the generalised repressilator. The repressilator is a ring of three mutually repressing genes pioneered in \cite{elowitz2000synthetic}, and theoretically generalised to rings consisting of more than three genes in \cite{Smith87}. Repression is an interaction between two genes, such that the protein product of the repressing gene prevents protein expression of the repressed gene. Or simply put, where one gene can turn off the other. A generalised repressilator with a sufficiently large, even number of genes (such as the four-gene ring in \figref{fig:gr}) can exhibit decaying but very long-lived oscillations~\cite{Strelkowa10}. The objective of this paper is to force one or several protein concentrations to follow \emph{a priori} defined reference trajectories, namely: sinusoids and ramps. A deeper background on synthetic biology and its control applications is provided in our 
companion paper \cite{sootla2013toggle}.

The paper is organised as follows. Section~\ref{sec:prel} covers batch-mode reinforcement learning. In Section~\ref{sec:track}, an extension of the Fitted $Q$ Iteration to the reference tracking problem is presented. In Section~\ref{sec:app} our illustrative application, reference tracking for the generalised repressilator system, is given. Finally, Section~\ref{sec:con} concludes the paper.

\section{Fitted $Q$ Iteration \label{sec:prel}}
\subsection{Problem Formulation}
Consider a deterministic discrete-time dynamical system
\begin{equation}
 \begin{aligned}
  \bfn_{t+1} = f(\bfn_t, \bfu_t)
 \end{aligned}\label{eq:sys}
\end{equation}
where the bold values $\boldsymbol n$ stand for vectors with elements $n^i$, $\bfn_t$ is the state of the system at time $t$, and $\bfu_t$ is a control input, which at each time $t$ belongs to a compact set $\setU$. Consider also, associated with this dynamical system, an optimal control problem, which is defined in terms of the minimisation of an infinite sum of discounted costs $\bfc(\bfn, \bfu)$:
\[
\min_{\mu(\cdot)}\sum_{i=t}^{\infty} \gamma^{i-t} \bfc(\bfn_i, \mu(\bfn_i))
\]
where $\mu(\cdot)$ is a mapping from the state-space onto $\setU$, which is called the feedback control policy, and $\gamma$ is a positive constant smaller than one, which is called the discount factor. For the purpose of this paper, it is assumed that $\bfc(\cdot,\cdot)$ is a given function. The goal is to compute the optimal policy based only on one-step transitions of the system \eqref{eq:sys}. One-step system transitions are given as a set $\calF = \{\bfn_l, \bfu_l, \bfn_l^+\}_{l=1}^{\#\calF}$, where $\bfn_l^+$ denotes a successor state of the system in state $\bfn_l$ subjected to input $\bfu_l$ (if the function $f(\cdot,\cdot)$ is known then $\bfn_l^+$ is simply equal to $f(\bfn_l,\bfu_l)$). 

\subsection{Algorithm \label{ss:fittedQ}}
The central object in Fitted $Q$ Iteration is the $Q$ function, which is defined as follows:
\[
Q(\bfn_t, \bfu_t) = \bfc(\bfn_t, \bfu_t) +  \min_{\mu(\cdot)}\sum_{i=t+1}^{\infty} \gamma^{i-t} c(\bfn_i, \mu(\bfn_i)).
\]
The main idea of the approach is to exploit the celebrated Bellman equation 
\begin{equation} \label{eq:bellman}
Q(\bfn, \bfu) = \bfc(\bfn, \bfu) + \gamma \min_{\bfu'\in \setU}Q(f(\bfn,\bfu),\bfu'),
\end{equation}
which provides a convenient expression for the computation of the optimal feedback control policy:
\begin{equation}
{\mu}^{\ast}(\bfn) = \argmin_{\bfu \in \setU} Q(\bfn, \bfu)\label{eq:policy}
\end{equation}
The Bellman equation can be theoretically solved using an iterative procedure
\[
Q_k(\bfn, \bfu) = \bfc(\bfn, \bfu) + \gamma \min_{\bfu'\in \setU}Q_{k-1}(f(\bfn,\bfu),\bfu'),
\]
where $Q_0 = \bfc$ and $Q_{\infty}$ is the unique solution to \eqref{eq:bellman} due to the fact that $T(Q) = \bfc + \gamma \min_{\bfu'\in \setU} Q$ is a contraction mapping (cf. \cite{busoniu2010reinforcement}). However in the continuous state-space case, this iterative procedure is hard to solve, especially when only the one-step system transitions in $\calF$ are given. The Fitted Q iteration algorithm computes, from the sole knowledge of ${\mathcal F}$ a sequence of functions $\hat{Q}_1$, $\hat{Q}_2$, $\ldots$ that approximates the sequence $Q_1$, $Q_2$, $\ldots$. Let $\hatQ_0 = \bfc$  and for every $(\bfn_l,\bfu_l,\bfn_l^+)$ in $\calF$ compute:
\begin{equation}
\label{eq:fitQ1}
\hatQ_{1}(\bfn_l, \bfu_l) = \bfc(\bfn_l, \bfu_l) + \gamma \min_{\bfu\in \setU}\hatQ_{0}(\bfn_l^+, \bfu).
\end{equation}
This expression gives $\hatQ_1$ only for $\bfn_l$, $\bfu_l$ in $\calF$, while the entire function $\hatQ_1(\cdot,\cdot)$ is estimated using a regression algorithm (\eg EXTRA Trees \cite{Geurts:2006}). Now the iterative procedure is derived by generalising \eqref{eq:fitQ1} as follows:
\begin{equation}
\label{eq:fitQ}
\hatQ_{k}(\bfn_l, \bfu_l) = \bfc(\bfn_l, \bfu_l) + \gamma \min_{\bfu\in \setU}\hatQ_{k-1}(\bfn_l^+, \bfu),
\end{equation}
where at each step $\hatQ_k(\cdot,\cdot)$ is estimated using a regression algorithm.
If the iteration procedure stops at the iteration number $k$, an approximate policy can be computed as follows:
\begin{equation}
\hat{\mu}^{\ast}(\bfn) = \argmin_{\bfu \in \setU} \hatQ_{k}(\bfn, \bfu)\label{eq:policy_fqi}
\end{equation}
\begin{algorithm}[t]
\caption{Fitted $Q$ iteration\label{alg:fqi}}
{\bf Inputs:} Set of one-step system transitions $\calF = \{\bfn_l, \bfu_l, \bfn_l^+\}_{l=1}^{\#\calF}$, cost $\bfc(\cdot,\cdot)$, stopping criterion\\
{\bf Outputs:} Policy $\hat\mu^{\ast}(\bfn)$
\begin{algorithmic}
\STATE $k \leftarrow 0$  
\STATE $\hatQ_0(\cdot,\cdot)\leftarrow \bfc(\cdot,\cdot)$
\REPEAT
\STATE $k \leftarrow k+1$ 
\STATE Compute \eqref{eq:fitQ} to obtain the values of $\hatQ_k(\cdot, \cdot)$ for all $\{\bfn_l, \bfu_l\}$ in $\calF$
\STATE Estimate the function $\hatQ_k(\bfn,\bfu)$ using a regression algorithm with input pairs $(\bfn_l, \bfu_l)$ and function values $\hatQ_k(\bfn_l, \bfu_l)$.
\UNTIL{stopping criterion is satisfied}
\STATE Compute the policy according to \eqref{eq:policy_fqi}
\end{algorithmic}
\end{algorithm}  
A simple stopping criterion can be, for example, an \emph{a priori} fixed maximum number of iterations $N_{\rm it}$. The value $N_{\rm it}$ is chosen such that $\gamma^{N_{\rm it}}$ is sufficiently small and, hence, the values $\hatQ_k(\bfn_l, \bfu_l)$ are not modified significantly for $k$ larger than $N_{\rm it}$. Other, more advanced, stopping criteria are described in \cite{Ernst2005c}. The resulting iterative method is outlined in Algorithm \ref{alg:fqi}. A major property of the fitted $Q$ iteration algorithm is convergence. This is understood as convergence of $\hatQ_k$ to a fixed state-action value function $\hatQ^{\ast}$ given a fixed set $\calF$ as $k$ grows to infinity. It was shown in \cite{Ernst2005c}, that under certain conditions Algorithm~\ref{alg:fqi} converges and it is possible to estimate in advance the distance between $\hatQ^{\ast}$ and the iterate $\hatQ_k$.

\section{Periodic Reference Tracking Using the Fitted $Q$ Iteration Algorithm \label{sec:track}}
\subsection{Problem Formulation\label{ss:pf}}
Consider, a system:
\begin{equation}
\begin{aligned}
 \bfn_{t+1} &= f(\bfn_t, \bfu_t)\\
 \bfr_{t+1} &= g(\bfr_t) 
\end{aligned} \label{eq:sys_ext}
\end{equation}
where the function $f(\cdot,\cdot)$ is unknown. The function $g(\cdot)$, however, is a known periodic function of period $T$, and the variable $\bfr$ takes only a finite number of values $\{\bfv_i\}_{i=1}^T$. Hence, $g(\bfv_{i})$ is equal to $\bfv_{i+1}$ for all $i$ smaller than $T$, and $g(\bfv_T)$ is equal to $\bfv_1$. The reference tracking problem is defined as follows: 
\begin{equation}
\begin{aligned}
\min_{\mu(\cdot,\cdot)}&\sum_{i=t}^{\infty} \gamma^{i-t} \bfc( d(\bfn_i, \bfr_i), \bfn_i, \bfu_i)  \\
\textrm{ subject to }& \textrm{ system dynamics }\eqref{eq:sys_ext}, \textrm{ and }  \\
                     & \mu(\bfn_t,\bfr_t) = \bfu_t\in\setU
\end{aligned}\label{min:track}
\end{equation}
where $\bfc$ is a given instantaneous cost function, $d(\cdot,\cdot)$ is a function defining the distance between the current state $\bfn_i$ and reference $\bfr_i$, and $\mu(\cdot,\cdot)$ is a feedback control policy. In order to track the reference $\bfr$, reducing the value of the cost $\bfc$ must reduce the distance $d$. The instantaneous cost can optionally depend on the control signal $\bfu$ and the states $\bfn$ in order to provide additional constraints in the state-space and/or control action space. As for the Fitted $Q$ Iteration, the control policy is inferred based solely on the trajectories given in the form of one-step system transitions $\calF = \{\bfn_l, \bfu_l, \bfn_l^+ \}_{l=1}^{\#\calF}$, where $\bfn_l^+$ denotes a successor state of the system in state $\bfn_l$ subjected to input $\bfu_l$. 
\subsection{Algorithm}
The variable $\bfr$ can be seen as an additional state in the extended state-space $\{\bfn, \bfr \}$. Based on this extended state-space we can derive the Bellman equation for the tracking problem and the following iterative procedure for the computation of the $Q$ function:
\begin{multline}
 Q_{k}(\bfn, \bfr, \bfu) = \bfc(d(\bfn, \bfr), \bfn, \bfu) + \\
 \min\limits_{\bfu'\in \setU} Q_{k-1}(\bfn^+, g(\bfr), \bfu')\quad \forall \bfn, \bfr, \bfu ~\label{eq:fqi_e}
\end{multline}
where $Q_0$ is equal to $\bfc$. It can be shown that this iterative procedure has a unique solution $Q^{\ast}$ based on a similar contraction mapping argument as in the previous section. Hence the optimal policy in the extended state-space is computed as:
\[
 \mu(\bfn, \bfr) = \min\limits_{\bfu\in \setU} Q^{\ast}(\bfn, \bfr, \bfu)
\]
The input set to the algorithm normally consists of the one-step system transitions $\calF = \{\bfn_l, \bfu_l, \bfn_l^+ \}_{l=1}^{\#\calF}$. However, since our state-space has been extended to include $\bfr$, $\calF$ should now include $\bfr$ as well. Since the time evolution of $\bfr$ is known \emph{a priori} we can simply modify the set as follows $\calTF = \{\bfn_l, \bfv_i, \bfu_l, \bfn_l^+, g(\bfv_i)\}_{i,l}$, where $i=1,\dots,T$ and $l=1,\dots,\#\calF$. This will effectively copy $T$ times the training set $\calF$. Now given this modification, we can proceed to the computational procedure. As before $\hatQ_0$ is equal to $\bfc$ and the next iterates can be obtained according to:
\begin{multline}
 \hatQ_{k}(\bfn_l, \bfv_i, \bfu_l) = \bfc(d(\bfn_l, \bfv_i), \bfn_l, \bfu_l) + \\
 \min\limits_{\bfu'\in \setU} \hatQ_{k-1}(\bfn_l^+, g(\bfv_i), \bfu') \quad \forall l,\, \bfv_i~\label{eq:fqi_e_app}
\end{multline}
According to the Fitted Q Iteration framework, every function $\hatQ_{k}(\cdot,\cdot,\cdot)$ must be estimated by a regression algorithm, which uses the input set $\{\bfn_l, \bfv_i, \bfu_l\}_{i,l}$ and the corresponding values of the approximated function $\hatQ_{k}(\bfn_l, \bfv_i, \bfu_l)$. This input set can grow significantly, if the period $T$ of the reference trajectory is large, which can render a regression algorithm computationally intractable. For example, with only a thousand one-step system transitions $\{\bfn_l, \bfu_l, \bfn_l^+ \}_{l=1}^{\#\calF}$ and period $T$ equal to $200$, the total number of samples $\{\bfn_l, \bfv_i, \bfu_l\}_{i,l}$ is equal to $200\,000$. 
Therefore, it is proposed to break up the regression of $\hatQ_k(\cdot,\cdot,\cdot)$ into $T$ independent regression problems, one for every function $\hatQ_{k}(\cdot, \bfv_i, \cdot)$.
This can be done, because the evolution of the variable $\bfr$ is known in advance and $\bfr$ takes a finite number of values. To make these ideas more transparent, \eqref{eq:fqi_e_app} is rewritten using a different notation as follows:
\begin{multline}
 \hatQ^{\bfv_i}_{k}(\bfn_l, \bfu_l) = \bfc(d(\bfn_l, \bfv_i), \bfn_l, \bfu_l) + \\ \min\limits_{\bfu\in \setU} \hatQ^{g(\bfv_{i})}_{k-1}(\bfn_l^+, \bfu) \quad \forall l,\, \bfv_i,~\label{eq:fqi_ref}
\end{multline}
where $\hatQ^{\bfv_i}_{k}(\cdot, \cdot)$ stands for the function $\hatQ_k(\cdot, \bfv_i, \cdot)$. For every value $\bfv_i$, the regression algorithm will approximate the function $\hatQ_{k}^{\bfv_i}(\cdot, \cdot)$ by 
using the input set $\{\bfn_l, \bfu_l\}_{l}$ and the corresponding values $\hatQ_{k}^{\bfv_i}(\bfn_l, \bfu_l)$. Thus the initial regression problem is indeed separated into $T$ independent problems.

Our approach can be also seen as a modification of the $\hatQ_k$ function approximator. The first layer of our approximator is a deterministic branching according to the values $\bfv_i$. After that a regression algorithm is performed to approximate the functions $\hatQ_k^{\bfv_i}(\cdot, \cdot)$ as prescribed. Finally, if the iterative procedure has stopped at iteration $k$, a near-optimal policy can be computed as follows:
\begin{equation}\label{eq:pol}
 \hat\mu^{\ast}(\bfn, \bfr) = \min\limits_{\bfu\in \setU} \hatQ_k^{\bfr}(\bfn, \bfu)
\end{equation}

\begin{algorithm}[t]
\caption{Reference Tracking Fitted $Q$ Iteration \label{alg:rtfqi}}
{\bf Inputs:} Sets of one-step system transitions $\calF = \{\bfn_l, \bfu_l, \bfn_l^+\}_{l=1}^{\#\calF}$, function $g(\cdot)$ and reference values $\{\bfv_i\}_{i=1}^T$, cost $\bfc(d(\cdot,\cdot),\cdot,\cdot)$, stopping  criterion\\
{\bf Outputs:} Policy $\hat\mu^*(\bfn,\bfr)$
\begin{algorithmic}
\STATE $k \leftarrow 0$ 

\STATE $\hatQ_0^{\cdot}(\cdot,\cdot)\leftarrow \bfc(d(\cdot,\cdot),\cdot,\cdot)$
\REPEAT
\STATE $k \leftarrow k+1$ 
\STATE Compute \eqref{eq:fqi_ref} to obtain the values of $\hatQ_k^{\bfv_i}(\cdot, \cdot)$ for all $\{\bfn_l, \bfu_l\}$ in $\calF$ and $\bfv_i$
\STATE Estimate the functions $\hatQ_k^{\bfv_i}(\bfn,\bfu)$ for every $\bfv_i$ using a regression algorithm 
with input pairs $(\bfn_l, \bfu_l)$ and function values $\hatQ_k^{\bfv_i}(\bfn_l, \bfu_l)$.
\UNTIL{stopping criterion is satisfied}
\STATE Compute the policy using \eqref{eq:pol}
\end{algorithmic}
\end{algorithm}  

Our approach is outlined in Algorithm~\ref{alg:rtfqi}. Periodicity is a crucial assumption, since the period of the reference trajectory corresponds to the number of different $\hatQ^{\bfv}_k$ functions built in this algorithm. Convergence of Algorithm~\ref{alg:rtfqi} can be established using the following lemma.
\begin{lem}
 Algorithm~\ref{alg:rtfqi} converges under similar conditions and considerations as the fitted $Q$ iteration algorithm in~\cite{Ernst2005c}. Moreover, the stopping criteria from~\cite{Ernst2005c} can be directly applied to Algorithm~\ref{alg:rtfqi}.
\end{lem}
To prove this lemma, we have to make sure that the approximator of the $\hatQ_k$ functions will not break the convergence proof in~\cite{Ernst2005c}. Only one modification in this approximator is made, which is the deterministic branching in its first layer according to $\bfv_i$. It can be shown that this modification does not violate the convergence arguments in~\cite{Ernst2005c}. 

Note, if we assume that the cost function is time-independent,  \ie $\bfr$ is constant and equal to $\bfv$, and $g(\bfv)$ is equal to $\bfv$, equation \eqref{eq:fqi_ref} becomes:
\begin{multline*}
 \hatQ^{\bfv}_{k}(\bfn_l, \bfu_l) = \bfc(d(\bfn_l, \bfv), \bfn_l, \bfu_l) + \\
 \min\limits_{\bfu'\in \setU} \hatQ^{\bfv}_{k-1}(\bfn_l^+, \bfu')\quad\forall l
\end{multline*}
which reduces Algorithm~\ref{alg:rtfqi} to Algorithm~\ref{alg:fqi} without any artefacts. Algorithm~\ref{alg:fqi} is applied to control of a gene regulatory network in our companion paper \cite{sootla2013toggle}.

\subsection{Other Applications of the Algorithm}
The presented algorithm addresses the problem of optimal control of the system \eqref{eq:sys_ext}, where a part of the dynamics is known and a part of the dynamics is to be learned. In our setting, the known dynamics describe a reference trajectory, which takes only a finite number of values. However, there are other problems for which Algorithm \ref{alg:rtfqi} can be useful. Consider the system, where $f(\cdot,\cdot)$ describes the system dynamics and $g(\cdot,\cdot)$ is known:
\begin{equation}
\begin{aligned}
 \bfn^{+} &= f(\bfn, \bfr)\\
 \bfr^{+} &= g(\bfr, \bfu)
\end{aligned} \label{eq:sys_del}
\end{equation}
The function $g(\cdot,\cdot)$ could be a model of an actuator or some dynamics of the system, \eg a time-delay in the control signal. In the latter case, the function $g(\cdot, \cdot)$ delays the application of the control action by a number of time samples. If $\bfu$ takes a finite number of values, then the problem can be solved using Algorithm \ref{alg:rtfqi} without major modifications.

Another application is a system with dynamics $f(\cdot, \cdot)$ and a given feedforward compensator $g(\cdot,\cdot)$, which helps to counteract the measured disturbance $\bfd$:
\begin{equation}
\begin{aligned}
 \bfn^{+} &= f(\bfn, \bfu + \bfr)\\
 \bfr^{+} &= g(\bfr, \bfd)
\end{aligned} \label{eq:sys_ff}
\end{equation}
\begin{figure}[t]
\centering
\includegraphics[width=0.8\columnwidth]{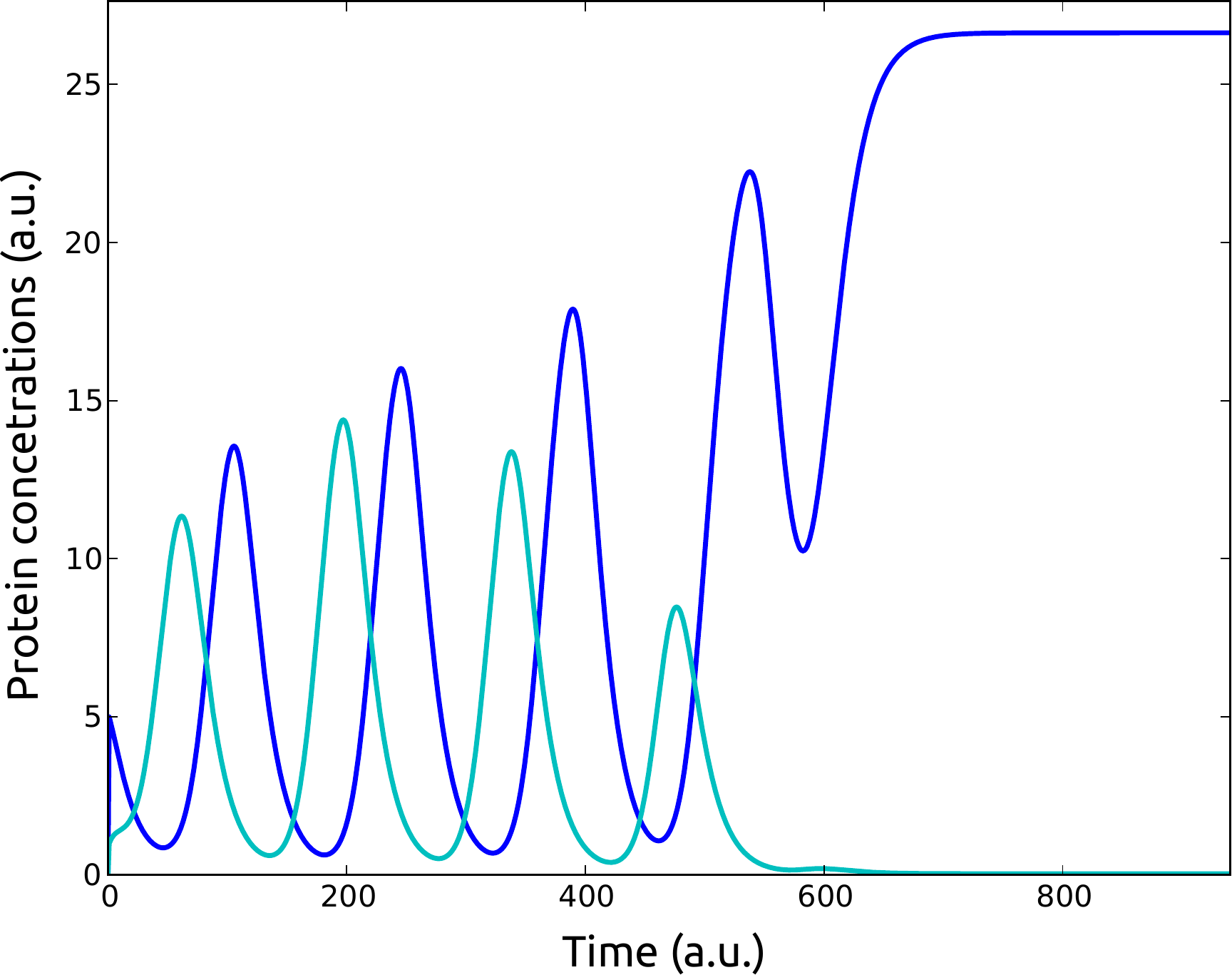}
\caption{Natural oscillatory behaviour of a generalised repressilator system consisting of $6$ genes. The blue line represents the time evolution of the protein concentration produced by the expression of gene $1$; the cyan line represents the time evolution of the protein concentration produced by the expression of gene $2$. The oscillations have a period of approximately $150$ arbitrary time units but are, however, not stable.} \label{fig:gr_nc}
\end{figure}

\section{Tracking of Periodic Trajectories by the Generalized Repressilator \label{sec:app}}
\subsection{System Description}
As an illustrative application of our method we consider the problem of periodic reference tracking for a six gene generalised repressilator system. There are two major species associated with every gene (the mRNA and protein concentrations), which results in a twelve state system. Throughout the section we adopt the following notation: $p^i$ denotes the protein concentration produced by the translation of mRNA $m^i$ of gene $i$. By definition of the generalised repressilator, the transcription of mRNA $m^i$ is repressed by the previous gene expression product $p^{i-1}$ in the network. With a slight abuse of notation, we assume that $p^{-1}$ is equal to $p^n$ in order to model the cyclic structure of the generalised repressilator. The dynamics of the generalised repressilator system can be described by the following set of deterministic equations:
\begin{equation}
\begin{aligned}
\dot{m^i}&=\frac{c_{1}^i}{1+(p^{i-1})^2}-c_{2}^i m^i  + \delta_{i1} b_1 u^1  + \delta_{i2} b_2 u^2 & \\
\dot{p^i}&=c_{3}^i m^i-c_{4}^i p^i&
\end{aligned} 
\label{eq:gr_det}
\end{equation}
where $i$ is an integer from one to six, and $\delta_{ij}$ is equal to one, if $i$ is equal to $j$, and equal to zero otherwise. We consider a control scheme similar to the one used for the toggle switch regulation problem (see our companion paper \cite{sootla2013toggle}), where the light control signals induce the expression of genes through the activation of their photo-sensitive promoters. The control signal $u^1$ only acts on the mRNA dynamics of gene~$1$, whereas the control signal $u^2$ only acts on the mRNA dynamics of gene~$2$. The influence of the light signals on the rate of mRNA production of genes~$1$ and $2$ is denoted by $b_1$ and $b_2$, respectively. To simplify the system dynamics and as it is usually done for the repressilator model \cite{elowitz2000synthetic}, we consider the corresponding parameters of the mRNA and protein dynamics for different genes to be equal. Hence, the trajectories will be very similar between the different genes. For the purpose of this paper, we chose:
\[
\begin{aligned}
 \forall i: \,\, &  c_{1}^i = 1.6, & &c_{2}^i = 0.16, & &c_{3}^i = 0.16, &c_{4}^i = 0.06,\\
 & b_1 = b_2 = 5
\end{aligned}
\] 
As shown in \cite{Strelkowa10}, this system exhibits a long-lived oscillatory behaviour around an unstable limit cycle as depicted in \figref{fig:gr_nc}. The ``natural'' period of these slowly decaying oscillations is around $150$ arbitrary time units. 
\subsection{Algorithm Parameters and Implementation}
The instantaneous cost function is defined differently based on the considered example. In the first case, the objective is for the concentration of protein $p^2$ to track an \emph{a priori} specified reference trajectory. In the other cases, the objective is for the concentrations of two proteins $p^1$ and $p^2$ to track their respective reference trajectories. When tracking of two protein concentrations needs to be ensured, the cost function will depend on these two protein concentrations $p^1$ and $p^2$ and on the corresponding references $r^1_t$ and $r^2_t$. Note that here and in the sequel $r^i_t$ stands for the $i$-th element of the vector $\bfr_t$. When tracking needs to be ensured for only one protein concentration (i.e., $p^2$), the cost function will only depend on $p^2$ and the scalar reference $r_t^2$. In both cases, the cost depends on  the control signals $u^i$ in order to penalise the use of light stimuli and thus minimise the metabolic burden caused by heterologous gene expression. 
The cost is defined using a distance between the observed state $p^i$ and the reference trajectories $r^i_t$:
\begin{multline*}
\bfc(\boldsymbol p, \bfr, \bfu) = 100\cdot \alpha (p^1 - r^1_t)^2 + 100 \cdot (p^2 - r^2_t)^2 +\\ 0.05\cdot u^1+0.05 \cdot u^2
\end{multline*}
where $\alpha$ will be equal to one or zero depending on how many reference trajectories we want to track simultaneously. 

The discount factor $\gamma$ is equal to $0.75$, the choice of which is guided by considerations similar to the ones in~\cite{Ernst2005c}. The stopping criterion is simply a bound on the number of iterations, which for the purpose of this paper is $30$. The set of system transitions is generated according to the following procedure. $300$ system trajectories starting in a random initial state are generated. The control actions applied to generate the trajectories are random as well. Every trajectory has at most $300$ samples. These state transitions are then gathered in the set $\calF$. At every step every $Q^{\bfr_i}$ function is approximated using the regression algorithm EXTremly RAndomized Trees (EXTRA Trees), which was shown to be an effective regression algorithm for the fitted $Q$ iteration framework~\cite{Geurts:2006}. The parameters for the algorithm are set to the default values from \cite{Ernst2005c}. 

The algorithm is implemented in Python using the machine learning \cite{pedregosa2011scikit}, parallelisation \cite{joblib}, graphics \cite{Hunter:2007} and scientific computation \cite{scipy} toolboxes. 
\subsection{Results}
\subsubsection{One sinusoidal reference trajectory, different periods} In this example, we are going to force the concentration of the protein $2$ to track a sinusoid with different periods. The parameter $\alpha$ is set to zero and the sinusoid is chosen to resemble the natural oscillations in terms of amplitude and mean value:
\[
 r^2_t = 8 + 7 \cdot \sin(T t / (2 \pi)).
\]
\begin{figure}[t]
\centering
\includegraphics[width=0.8\columnwidth]{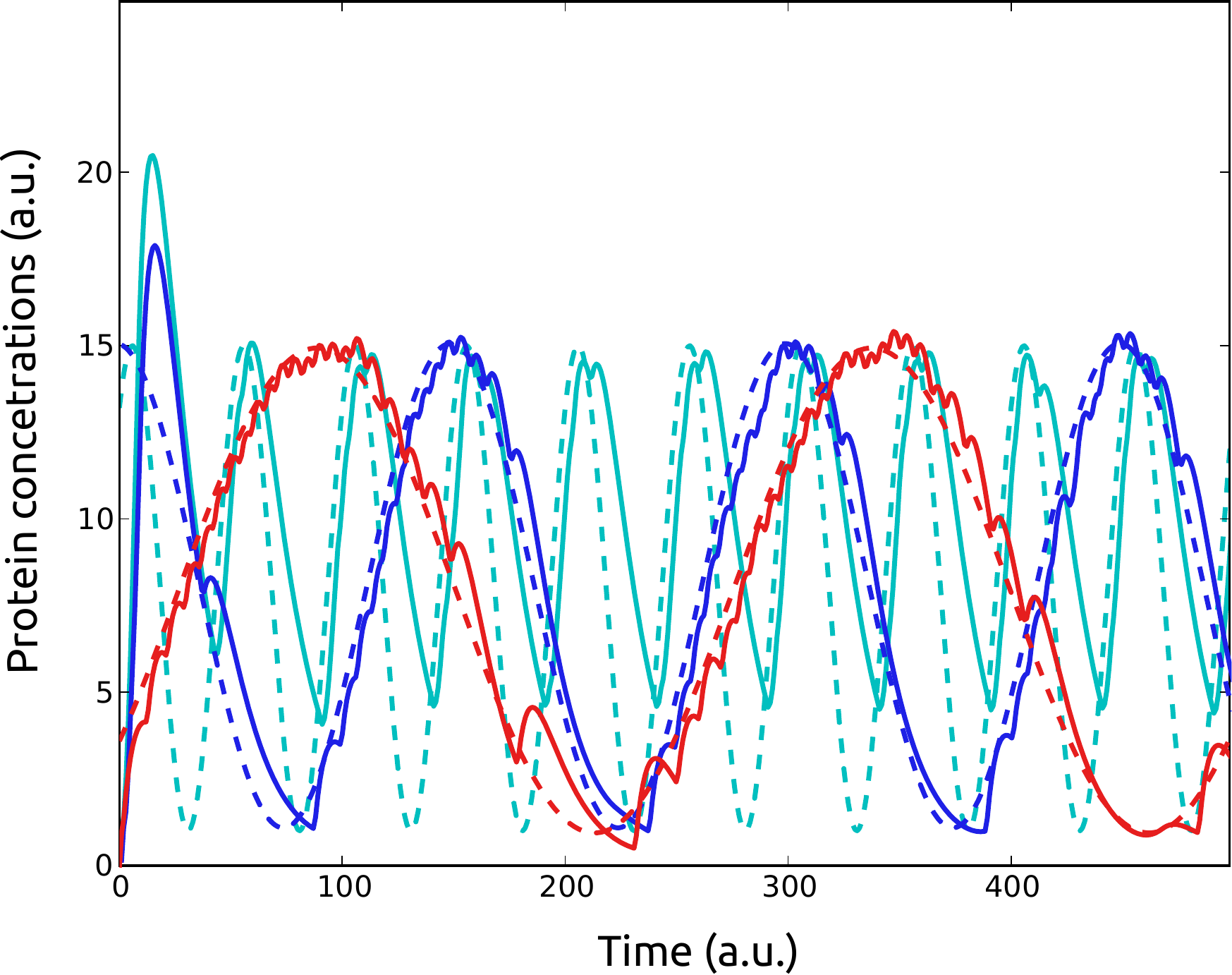}
\caption{Tracking a sinusoid in a six gene repressilator. The figure is colour coded: lines with the same colour correspond to the same experiment. Each solid curve represents the time evolution of the concentration protein $2$, which attempts to follow the same colour reference trajectory, represented by a dashed line. The cyan colour corresponds to the period $T=50$, the blue colour to $T=150$ and the red colour to $T=250$.}
\label{fig:gr_sin}
\end{figure}
We are testing the algorithm for the following periods $T=$ $50$, $150$, $250$. We can increase the concentration of the protein $2$ directly trough application of the control signal $u^2$. We can also decrease the concentration of the protein $2$ through increasing the concentration of the protein $1$, which acts as a repressor for gene $2$. The results of several experiments are depicted in \figref{fig:gr_sin}. The natural oscillations have a period of approximately $150$ arbitrary time units, therefore the blue dashed curve is easiest to follow. Using repression by gene $1$, the algorithm manages to find a schedule of light pulses that allows to track the dotted red curve, even though the period is much larger than $150$. However, due to the inherent dynamics of the system, the algorithm cannot bring down the concentration of protein $2$ fast enough to properly track the cyan curve. The repression by gene $1$ is not strong enough in this case to allow accurate tracking. It is important to remark that 
tracking of the trajectory by protein $2$ results in damping of the oscillations in the other proteins and mRNA dynamics. 

\subsubsection{Two sinusoidal reference trajectories} For this experiment we chose two sinusoids (\ie $\alpha$ is equal to one), where the second sinusoid lags behind the first one:
\[
\begin{aligned}
 r^1_t &= 8 + 7 \cdot \sin(200 t / (2 \pi))& \\
 r^2_t &= 8 + 7 \cdot \sin(200 (t + 200/3)/ (2 \pi))&
\end{aligned} 
\]
\begin{figure}[t]
\centering
\includegraphics[width=0.8\columnwidth]{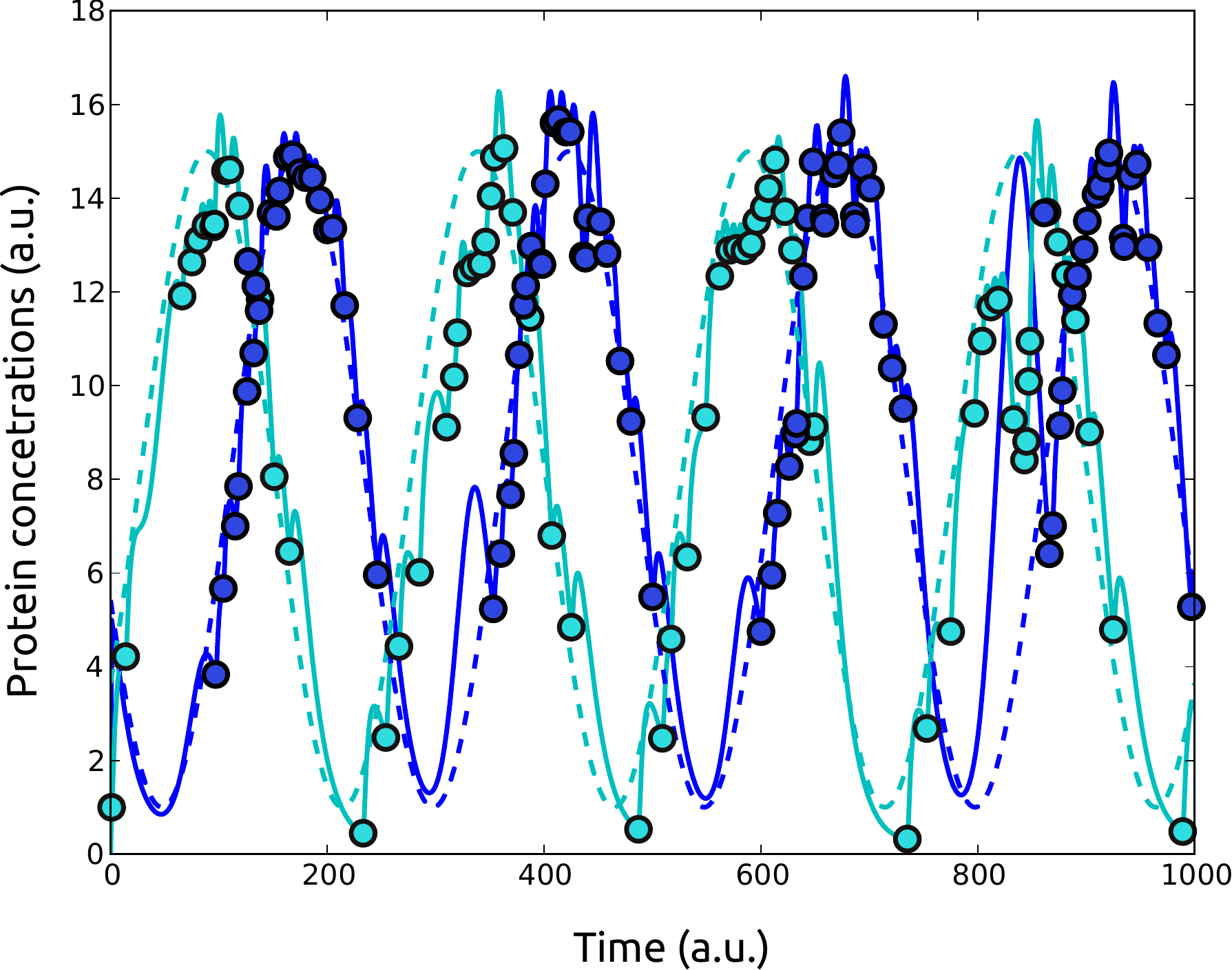}
\caption{Tracking two sinusoids in a six gene repressilator. The blue colour represents gene $1$ in the repressilator, which represses gene $2$ represented by the cyan colour. The dashed lines represent the reference trajectories; the solid lines represent the protein concentration tracking the reference with identical colour; finally, the coloured circled dots correspond to time instant at which control inputs in the form of light pulses were applied. Due to restrictions imposed by the system's dynamics the cyan reference should lag behind the blue one and the lag should be large enough to ensure appropriate tracking.
} \label{fig:gr_double_sin}
\end{figure}

\begin{figure}
\centering
\includegraphics[width=0.8\columnwidth]{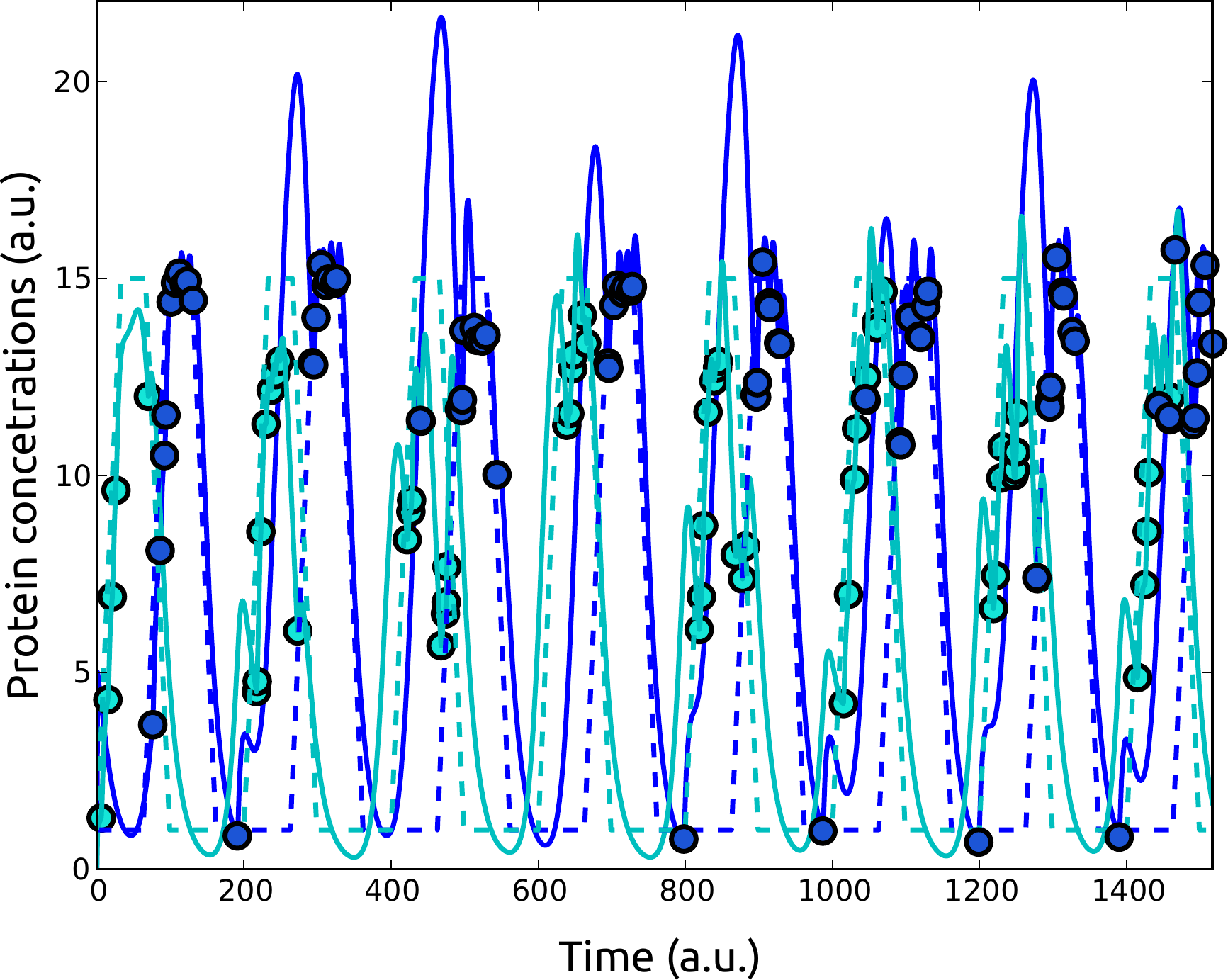}
\includegraphics[width=0.8\columnwidth]{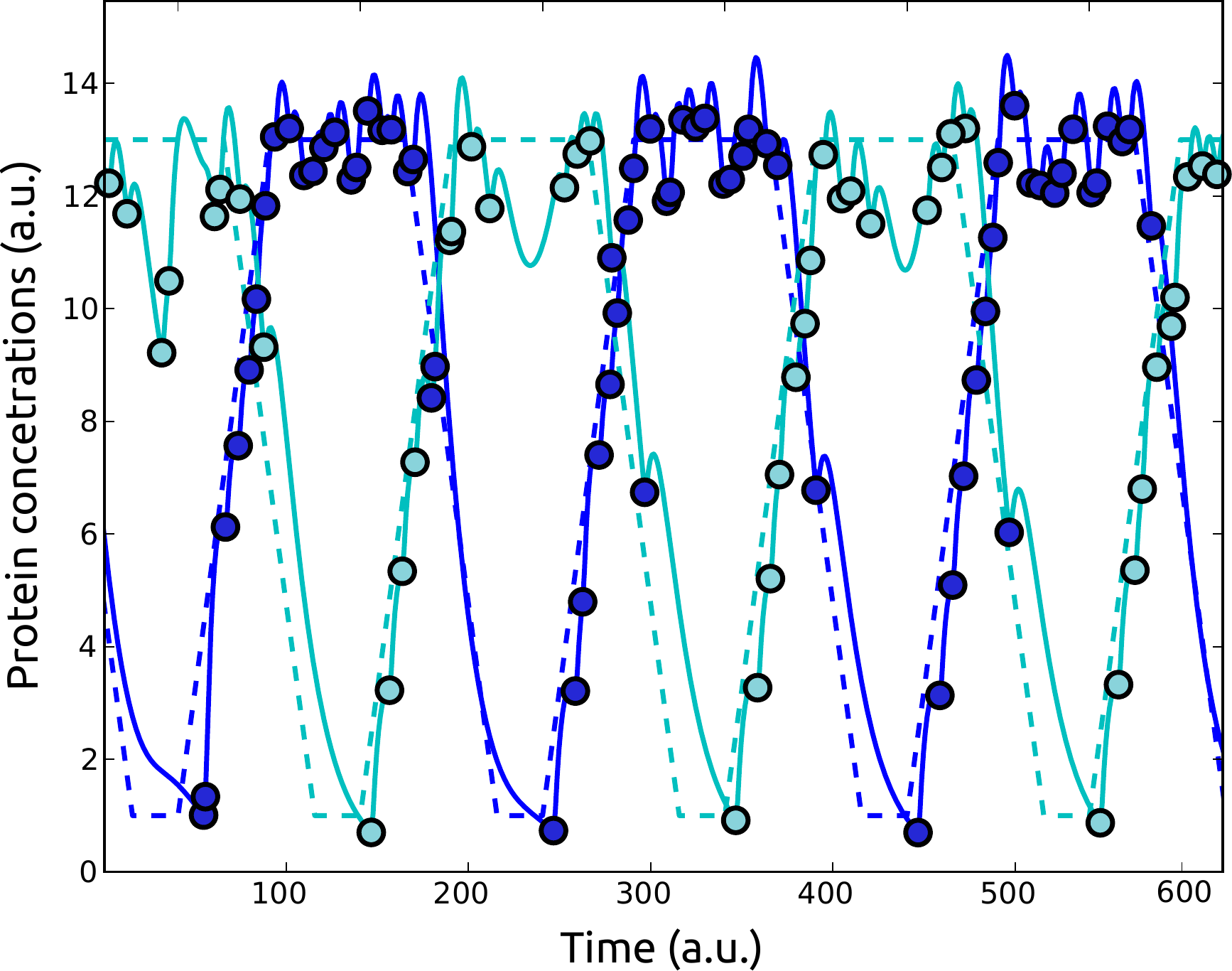}
\caption{Tracking the ramps in a six gene repressilator. Similar colour and line coding is used as in the figure above. In the upper panel, the algorithm finds it hard to keep both genes at low levels due to the inherent constraints imposed by the dynamics of the generalised repressilator; hence, the system cannot follow these ramp trajectories. Note that even though the first period is followed perfectly, the ``blue'' protein then starts to grow and we have no means to decrease its concentration. In the lower panel the ramp is changed so that the period of time spent at low concentrations is much shorter. This solves the above described issue.} \label{fig:gr_ramps}
\end{figure}
Genes, their protein products and the corresponding reference trajectories are colour coded in this example: the blue colour corresponds to protein $1$ and the cyan colour corresponds to protein $2$. The ``blue'' gene represses the ``cyan'' one, hence an increase in the ``blue'' protein concentration can be used to decrease the concentration of the ``cyan'' protein. However, the  concentration of the ``blue'' protein cannot be decreased directly. The sinusoids are chosen with approximate knowledge of the natural oscillation dynamics: in terms of amplitude and mean value of oscillations. Their period is chosen equal to $200$ time samples. As depicted in \figref{fig:gr_double_sin}, the algorithm can force the concentration of the first two proteins to follow both sinusoids. Moreover, numerous experiments were conducted starting from different initial conditions, which yielded similar tracking results after the initial transient period had elapsed. It is worth noting the interesting behaviour exhibited by the 
``blue'' protein concentration: At some point the protein concentration starts growing without any light stimulation. This can be explained by the repressilator oscillatory dynamics, where the protein's concentration can grow periodically. Moreover, since the ``blue'' protein concentration cannot be decreased directly it can grow significantly as can be observed during the time range between $800$ and $1000$ time units. It also means that controlling this system with a larger period will be harder due to this fast growth of the ``blue'' protein concentration induced by the dynamics of the system.

\subsubsection{Two ramp reference trajectories} The two ramp tracking setting is very similar to the two sinusoids tracking setting. However, in the situation depicted in upper panel of \figref{fig:gr_ramps} unsuccessful simultaneous tracking of two ramps is occurring. The reasons for such a behaviour are the same as above. The difference is that they are more pronounced in this case due to large time intervals during which a low reference value needs to be followed by one of the proteins. Note that the first ramp is followed almost perfectly by both protein concentrations. However, after a long time interval of low reference value, the concentration of the ``blue'' protein starts to grow due the inherent dynamics of the system. With a modified ramp such behaviour disappears as depicted in the lower panel of \figref{fig:gr_ramps}. Even though tracking is not perfect, the algorithm manages to find an approximate solution to the optimal control problem, which is quite remarkable given the very limited 
amount of information provided by the input-output data in such setting.

\section{Discussion and Conclusion\label{sec:con} }
In this paper, we have presented a periodic reference tracking reinforcement learning algorithm. The algorithm is based on the established fitted Q iteration framework, and inherits its properties. The proposed algorithm makes full use of the fact that the reference trajectory is known in advance, which results in better sample efficiency in comparison with other approaches proposed in the literature.

The algorithm is illustrated on the problem of tracking a periodic trajectory for the generalised repressilator system. This system has received considerable attention from the synthetic and systems biology community due to its ability to produce long-lived oscillatory behaviours. The algorithm was able to find near optimal solutions to the periodic tracking reference problem, even when the period of the reference trajectory was smaller than the natural period of oscillation of the system.

\bibliographystyle{hunsrt}
\bibliography{optimalControl}
\end{document}